\documentclass[aps,amsmath,amssymb,superscriptaddress,twocolumn,
prl,showpacs,showkeys,array]{revtex4}
\usepackage[dvips]{graphicx}
\usepackage{psfrag}

\newcommand{\s}{\sum\limits}

\newcommand{\be}{\begin{equation}}
\newcommand{\e}{\end{equation}}
\newcommand{\beml}{\begin{subequations}}
\newcommand{\eml}{\end{subequations}}
\newcommand{\beq}{\begin{eqnarray}}
\newcommand{\eq}{\end{eqnarray}}
\newcommand{\ba}{\begin{array}}
\newcommand{\ea}{\end{array}}
\newcommand{\lt}{\left}
\newcommand{\rt}{\right}
\newcommand{\n}{\nonumber}
\newcommand{\la}{\langle}
\newcommand{\ra}{\rangle}

\newcommand{\im}{\;{\rm Im}\,}

\newcommand{\ep}{\varepsilon}

\begin{document}

\author{M. E. Torio}
\affiliation{Instituto de F\'{\i}sica Rosario, CONICET-UNR, 
Bv. 27 de Febrero 210 bis, 2000 Rosario} 
\author{K. Hallberg}
\affiliation{Centro At\'omico Bariloche and Instituto Balseiro,
Comisi\'on Nacional de Energ\'{\i}a At\'omica, 8400 Bariloche, Argentina}
\author{S. Flach}
\author{A. E. Miroshnichenko}
\affiliation{Max-Planck-Institut f\"ur Physik komplexer Systeme, N\"othnitzer
Strasse 38, D-01187 Dresden, Germany}
\author{M. Titov}
\affiliation{Max-Planck-Institut f\"ur Physik komplexer Systeme, N\"othnitzer
Strasse 38, D-01187 Dresden, Germany}
\affiliation{Physics Department, Columbia University, New York, NY 10027, USA}

\title{Spin filters with Fano dots}
\date{\today}

\begin{abstract}
We compute the zero bias conductance of electrons through 
a single ballistic channel weakly coupled to a side 
quantum dot with Coulomb interaction.
In contrast to the standard setup which is designed 
to measure the transport through the dot, 
the channel conductance reveals Coulomb blockade 
dips rather then peaks due to the Fano-like backscattering.
At zero temperature the Kondo effect leads 
to the formation of broad valleys of small conductance 
corresponding to an odd number of electrons on the dot.
By applying a magnetic field in the dot region we find 
two dips corresponding to 
a total suppression in the conductance of spins up and down separated by an 
energy of the order of the Coulomb interaction. This provides a possibility 
of a 
perfect spin filter.
\end{abstract}


\maketitle

\section{Introduction}

In recent decades the electric transport through 
quantum dots (QD) has been extensively studied 
both theoretically and 
experimentally \cite{mesoscopic1,mesoscopic2,mesoscopic3}.
As the result a comprehensive picture of a big variety of
underlying physical phenomena has emerged 
(See e.g. \cite{ABGreview,Areview} and references therein).
Confinement of electrons in small quantum dots leads 
to the necessity of taking into account their Coulomb repulsion.
At finite temperatures the main effect
is the Coulomb blockade \cite{KMMTWW,AL}, 
which manifests itself in the 
oscillations of the dot's conductance versus the gate voltage
applied to the dot.
At very low temperature in the Coulomb blockade regime
the dot's conductance increases in so-called ``odd'' valleys,
where the number of electrons on the dot remains odd.
This is a signature of the mesoscopic Kondo effect \cite{KG,GR,NL},
which has been experimentally observed \cite{goldhaber98,cronenwet98,schmid98}.
In recent years, several authors have studied the possibility of spin 
polarization in transport through QD's \cite{costi,recher,nglee88,ohno} and other 
nanoscopic configurations like rings \cite{popp}, T-shaped spin filters 
\cite{kiselev} and dots connected to Luttinger liquid leads  
\cite{schmeltzer03}.
 In \cite{costi,recher,nglee88} the QD's are attached to two leads (right and
left), which we  call the {\it substitutional} configuration and the transport
properties are complementary to the case studied here.

In this report we combine the standard physics of the
Coulomb blockade with the additional interference
of the Fano type \cite{fano}. The latter
is essential only in the limit of 
a single conducting channel, which is 
weakly coupled to a side quantum dot \cite{goeres00,THCP}. 
The transport properties of such setup has been 
studied in the work by Kang {\it et al.} \cite{kang01}.
It was shown that the dependence of the channel conductance 
on the gate voltage applied to the dot is opposite
to the conductance of the dot itself.
Indeed the single-particle resonance 
locks the channel giving rise
to a resonant dip of zero conductance.
Therefore, at the Coulomb blockade regime
the channel conductance should reveal a sequence of sharp 
dips as the function of the gate voltage.
Similarly at very low temperature the Kondo effect
should lead to the appearance of a shallow 
and broad valley in the conductance versus 
the gate voltage curve, which 
corresponds to the odd number of electrons on the dot.

If the spin degeneracy is lifted, say, by 
an applied magnetic field, and the splitting 
exceeds the temperature, the resonances
for spin up and spin down electrons become well separated in the absence of 
the Kondo effect.
This gives rise to the resonance spin polarized
conductance due to the effect of Fano interference.

\section{The model}

In what follows we are dealing essentially
with the single-level Anderson impurity model 
\cite{hewson93}. 
We take advantage of recently studied side dot 
model \cite{kang01,THCP}
characterized by the one-dimensional Hamiltonian:
\beq
H&=&-t \s_{n,\sigma} 
(c^\dag_{n,\sigma} c_{n-1,\sigma}+c^\dag_{n,\sigma} c_{n+1,\sigma})
+\s_{\sigma}\ep_{0,\sigma} n_{\sigma}\n\\
&+&
\s_\sigma(V d^{\dag}_{\sigma}c_{0,\sigma}+V^\star c^{\dag}_{0,\sigma}d_{\sigma} )
+ U n_{\uparrow}n_{\downarrow}, 
\label{eq1}
\eq
where $n_{\sigma} = d^{\dagger}_{\sigma} d_{\sigma}$.
The $c$-operators describe the propagation of electrons
in the conducting channel, while
the $d$ operators describe the occupation of the side dot 
attached to site zero of the channel. The dot 
is modeled as an Anderson impurity with a Hubbard interaction
constant $U>0$ \cite{hewson93}. In this paper we consider 
$t=V=1$ and $\ep_F=0$ (half-filling) for all numerical computations.
We also assume that the local magnetic field
induces a Zeemann splitting of the dot levels
\be
\ep_{0,\uparrow}=\ep_0 + \Delta/2,\qquad
\ep_{0,\downarrow}=\ep_0 - \Delta/2,
\e
and the position of the resonant level $\ep_0$ 
can be adjusted by an external gate voltage.

The considered model falls into the class of Anderson 
Hamiltonians \cite{anderson61}. 
Let us emphasize that the conductance 
depends crucially on the geometry of the coupling
between the dot and the channel unlike
the thermodynamic properties of the model \cite{THCP},
such as the Kondo temperature $T_K$, etc. 
Below we are mostly concerned with the ground state properties
at zero temperature. 

\section{Fano resonance for $U=0$}

Let us consider first the case $U=0$. Then the 
dimensionless conductance per spin channel $g_\sigma$ is linked
to the transmission coefficient $\tau$ by the Landauer formula
\be
g_\sigma=|\tau_\sigma|^2.
\e
For $U=0$ the spin up and spin down electrons do not interact 
and we can suppress the index $\sigma$. 
The transmission can be computed within the
one-particle picture for an electron 
moving at the Fermi energy $\ep_F$:
\beml
\label{oneparticle}
\beq
-\ep_F \phi_i&=&t(\phi_{n-1}+\phi_{n+1})+V^\star\varphi\delta_{n0},\\
-\ep_F \varphi&=&-\ep_{0,\sigma}\varphi+V\phi_0,
\eq
\eml
where $\delta_{nm}$ is the Kronecker symbol,
$\phi_n$ refers to the amplitude 
of a single particle at site $n$ in the conducting channel
and $\varphi$ is the amplitude at the side dot.

The scattering problem is solved by the ansatz 
\be
\phi_{n<0}=e^{i q n}+\rho e^{-i q n},\qquad
\phi_{n>0}=\tau e^{i q n},\label{eq3}
\e
with the usual dispersion relation $\ep=-2 t \cos{q}$.  
Substituting Eq.~(\ref{eq3}) in Eq.~(\ref{oneparticle})
we readily find
\begin{equation}\label{eq4}
\tau=\frac{2 i t \sin{q}}{2 i t \sin{q}+|V|^2 (\ep_0-\ep_F)^{-1}},
\end{equation}
hence the conductance
\begin{equation}\label{eq9}
g_\sigma=|\tau_\sigma|^2=\left[1+\frac{\Gamma^2}{4(\ep_{0,\sigma}-\ep_F)^2}
\right]^{-1},
\end{equation}
where $\Gamma=2|V|^2/|v_F|$ with $v_F=d\ep/dq|\ep_F$ is the Fermi velocity. 

On the other hand the mean occupation number on 
the dot at zero temperature
$\left\langle n_{\sigma} \right\rangle$ 
is related to the scattering phase shift
by the Friedel sum rule \cite{langreth66,hewson93}
\begin{equation}
\left\langle n_{\sigma} \right\rangle=-\frac{1}{\pi}
\rm{Im}\; \ln{(\ep_F-\ep_{0,\sigma}+i\Gamma/2)}
\end{equation}
suggesting the relation
\begin{equation}
\label{cos}
g_\sigma=\cos^2\pi\left\langle n_{\sigma} \right\rangle.
\end{equation}
This relation has a geometric origin and actually 
holds for arbitrary $U$ provided zero temperature limit.
This is in contrast to Eqs.\ (\ref{eq4},\ref{eq9}) 
which are applicable only for $U=0$.

The symmetric form of the Fano resonance 
given by Eq.~(\ref{eq9}) is based on the symmetry
of the coupling term in the Hamiltonian (\ref{eq1}).
For nonzero magnetic field $\Delta \gg \Gamma$ 
the two Fano resonances for spin up and spin down 
electrons are energetically separated. Therefore,
the current through the channel is completely 
polarized at $\ep_F=\ep_{0,\uparrow}$ and
$\ep_F = \ep_{0,\downarrow}$.

\section{Fano resonance for $U \neq 0$ at the degenerate level $\Delta=0$}

Let us consider the case of zero magnetic field $\Delta=0$ 
but finite Coulomb interaction $U$. For zero coupling, $V=0$, 
the single particle excitations on the dot are 
located at energies $\ep_0$ and $\ep_0+U$. 
When the dot is weakly connected to the conducting channel 
(and at very low temperatures, $T<T_K$), 
a resonant peak in the density of states on the dot
arises at the Fermi energy due to the Kondo effect.
By using an approach identical to that of Refs.\ \cite{GR,NL},
we conclude that the channel conductance $g_{\sigma}$
is controlled entirely by the expectation value of the
particle number $\langle n_{\sigma} \rangle$ 
as given by Eq.~(\ref{cos}). This reasoning justifies  
Eq.~(\ref{cos}) for any values of $U$ and $\Delta$ \cite{kang-remark}.
Note that in the standard setup the dot's conductance 
is proportional to $\sin^2 (\pi \langle n_{\sigma} \rangle )$
\cite{langreth66,GR,NL} due to the different geometry of the 
underlying system.

If the resonant level is spin degenerate $\Delta=0$ one has
$\langle n_{\uparrow} \rangle = \langle n_{\downarrow} \rangle$
independent of the value of $U$. The mean occupation of the dot
$\langle n_{\sigma} \rangle$ is a monotonous function of $\ep_0$,
ranging from 1 for $\ep_0 \ll \ep_F$ to 0 for $\ep_0 \gg \ep_F+U$.
According to Eq.~(\ref{cos}) the channel conductance 
vanishes for $\lt\la n_\sigma \rt\ra=1/2$.

\begin{figure}[t]
\includegraphics[width=7.6cm]{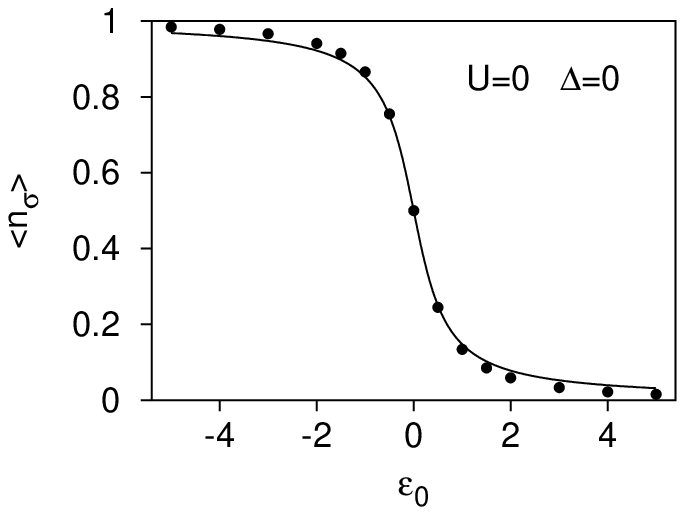}
\includegraphics[width=7.6cm]{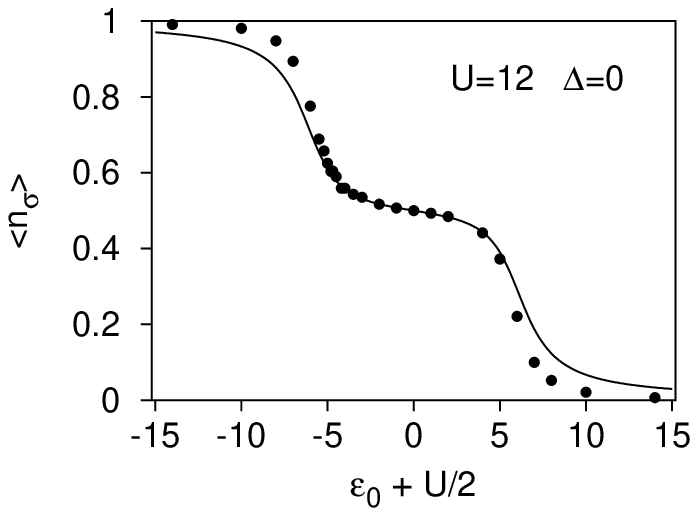}
\caption{The dependence of $\langle n_{\sigma} \rangle$
on the position of the resonance level $\ep_0$ 
for zero magnetic field, $U=0$ (top) and $U=12$ (bottom).
The results of the numerical calculation (black dots) 
are compared to the exact results by Wiegmann and 
Tsvelick \cite{WT} (solid line).}
\label{fig:n}
\end{figure}

\begin{figure}[t]
\includegraphics[width=7.6cm]{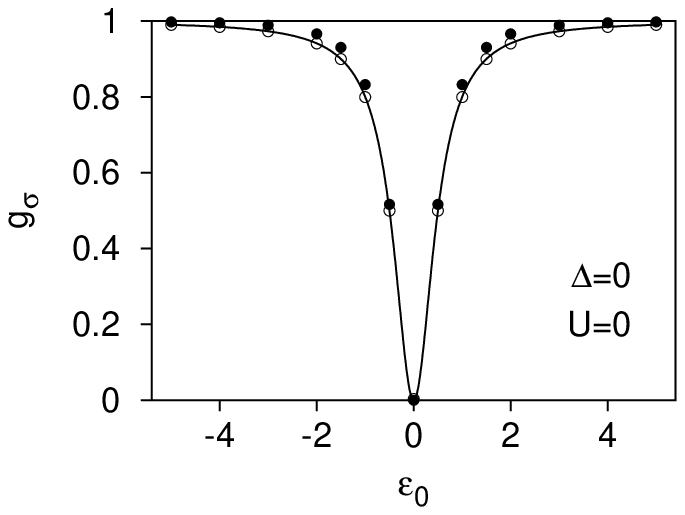}
\includegraphics[width=7.6cm]{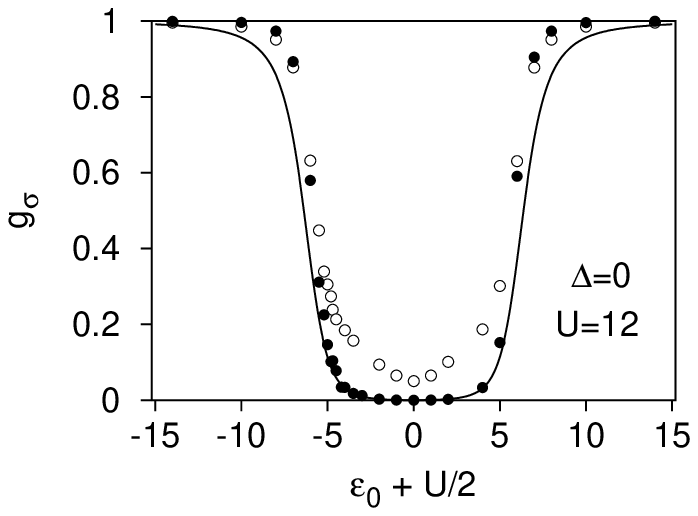}
\caption{The channel conductance $g_\sigma$
versus the resonance energy $\ep_0$ for zero magnetic field,
$U=0$ (top) and $U=12$ (bottom). 
The conductance computed numerically by different means 
from Eq.~(\ref{gsigma}) (white dots) as well as 
from Eq.~(\ref{cos}) (black dots) is compared to 
the conductance obtained from the exact solution by 
Wiegmann and Tsvelick \cite{WT} (solid line).
}
\label{fig:g}
\end{figure}

In order to compute
$\langle n_{\sigma} \rangle$ for finite $U$
we use the numerical techniques developed in 
Ref.\ \cite{THCP}.
We compare our results to those obtained from
the exact solution for $\la n_\sigma\ra$ 
by Wiegmann and Tsvelick \cite{WT}.
The latter holds only for the linearized spectrum
(which is not a serious restriction at half-filling)
and has a compact form only in the case 
of zero magnetic field. 

Numerically we obtain $\langle n_{\sigma} \rangle$
by integrating  the density of 
states at the dot over the energy. This quantity has been 
previously calculated using a combined method.  
In the first place we consider an open 
finite cluster of $N=9$ sites which includes the impurity. 
This cluster is diagonalized
using the Lanczos algorithm \cite{lanczos}. We then   
embed the cluster in an external reservoir of electrons, which   
fixes the Fermi energy of the system, attaching two semi-infinite leads to its
right and left \cite{ferrari}. This is done by calculating the one-particle   
Green's function $\hat{G}$ of the whole system within the chain approximation
of a cumulant expansion \cite{cumulant} for the dressed propagators. This
leads to the Dyson equation $\hat{G}=\hat{G}\hat{g}+\hat{T}\hat{G}$, where 
$\hat{g}$ is the cluster Green's function obtained by the Lanczos method.  
Following Ref.\ \cite{ferrari}, 
the charge fluctuations inside the cluster are taken into
account by writing $\hat{g}$ as a combination of $n$ and $n+1$ particles
with weights $1-p$ and $p$ respectively: 
$\hat{g}=(1-p)\hat{g}_{n}+p\hat{g}_{n+1}$. 
The total charge of the cluster and the weight $p$ 
are calculated by solving the following equations
\beml
\beq
Q_{c} &=&n(1-p)+(n+1)p,  \\
Q_{c} &=&-\frac{1}{\pi}\int_{-\infty}^{\ep_{F}}\s_{i}
\im G_{ii}(\ep)\, d\ep,  
\eq
\eml
where $i$ counts the cluster sites. Once convergence is reached,
the density of states is obtained from $\hat{G}$. 
One should stress, however, that the method is reliable only 
if the cluster size is larger then the Kondo length.

We also perform the alternative calculation of the conductance
evaluating the local density of states
$\rho_{\sigma}(\ep)$ at site $0$ \cite{meir}
\begin{equation}
\label{gsigma}
g_{\sigma}=2\pi t \rho_{\sigma}(\ep_F).  
\end{equation}  

We plot $\langle n_{\sigma} \rangle$ versus  
the resonance level energy $\ep_0$ for $U=0$
and $U=12$. In the presence of Coulomb repulsion
the mean occupancy of the dot $\langle n_{\sigma} \rangle$ 
has a flat region near the value $1/2$ due to the Kondo effect.
This corresponds to a valley of small channel conductance
(see Fig.\ \ref{fig:g}). Note that similar results
for the limit $U \rightarrow \infty$ have been recently obtained
by Bulka and Stefanski \cite{bulka01}.

The conductance found from
the numerical evaluation of $\langle n_{\sigma} \rangle$  
for finite $U$ is in better agreement with the exact results \cite{WT}
than the conductance computed from Eq.~(\ref{gsigma}).
Presumably, the discrepancy originates from the 
finite-size effects (the Kondo length is larger 
than the system size used in the numerical computation). 
The numerical calculation of $\langle n_{\sigma}\rangle$, 
which involves the energy integration, appears to be more precise 
then that of the local density $\rho_{\sigma}(\ep_F)$.

\section{Spin filter for $U\neq 0$ and $\Delta \neq 0$}

\begin{figure}[t]
\includegraphics[width=7.6cm]{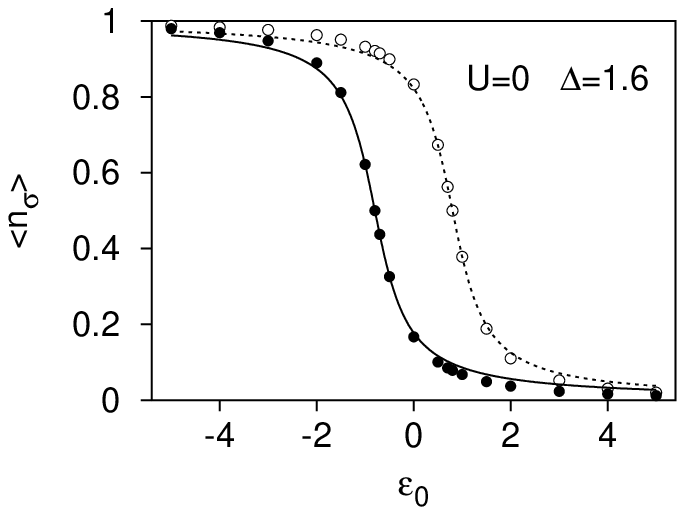}
\includegraphics[width=7.6cm]{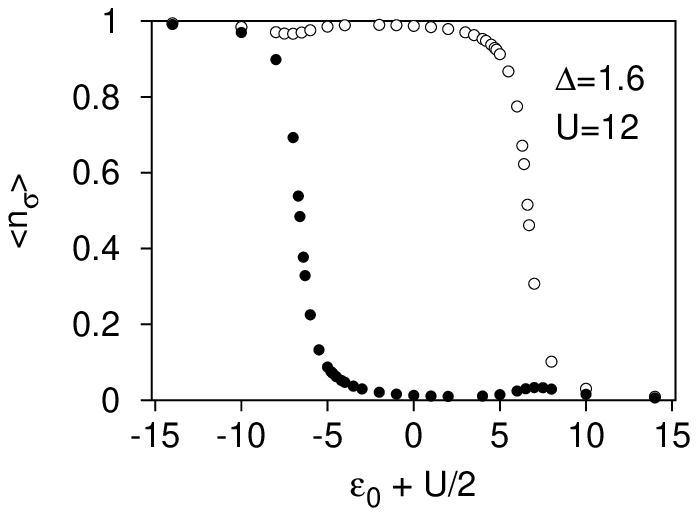}
\caption{The dependence of $\langle n_{\sigma} \rangle$
on the position of the resonance level $\ep_0$ 
for a finite splitting $\Delta=1.6$, $U=0$ (top) and $U=12$ (bottom).
The black (white) dots represent the numerical results for the
spin up (spin down) occupation number. The solid and the dashed 
line on the top figure represent the exact results of Eq. (\ref{eq9}).
}
\label{fig:nH}
\end{figure}

In the presence of a magnetic field 
the dot occupation for spin up and spin down electrons become
different. At zero temperature 
the conductance in each spin channel 
can be found from  Eq.~(\ref{cos}).
The expectation value $\lt\la n_\sigma\rt\ra$
decreases from $1$ to $0$ with increasing the 
resonant energy $\ep_0$, but it takes the value $1/2$ at different
gate voltages for $\sigma=\uparrow$ and $\sigma = \downarrow$ 
(see Fig.\ \ref{fig:nH}).

\begin{figure}[t]
\includegraphics[width=7.6cm]{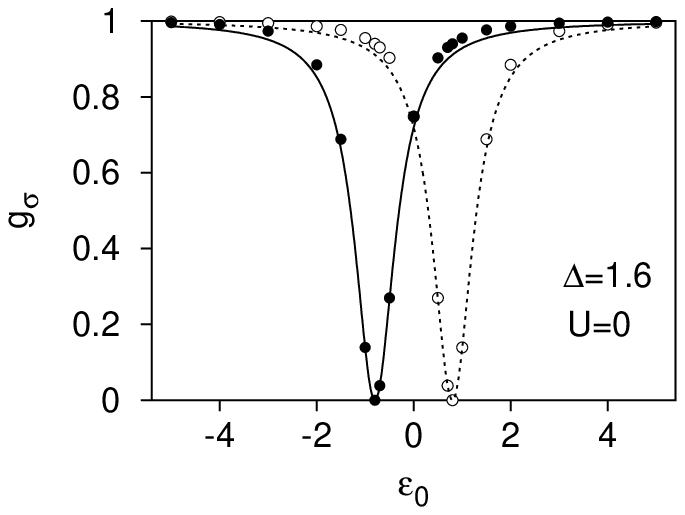}
\includegraphics[width=7.6cm]{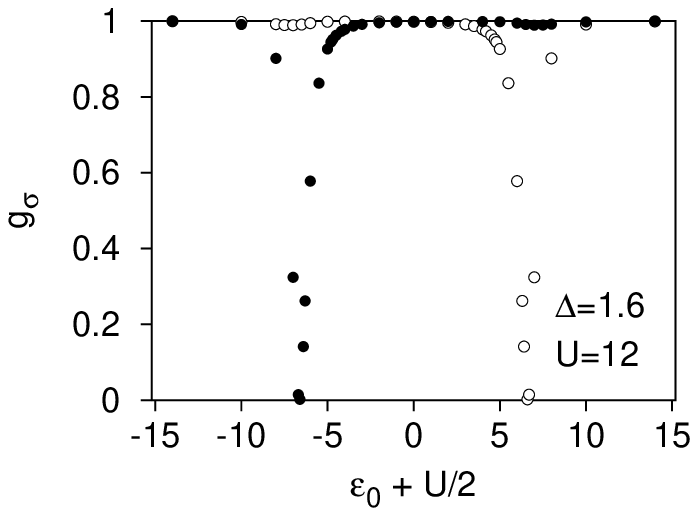}
\caption{The channel conductance $g_\sigma$
versus the resonance energy $\ep_0$ for 
finite splitting $\Delta=1.6$, $U=0$ (top) and $U=12$ (bottom). 
The conductance is computed numerically for spin up (black dots)
and spin down (white dots) electrons. The solid and the dashed 
line on the top figure represent the exact results of Eq. (\ref{eq9}).
}
\label{fig:gH}
\end{figure}

Consequently we may obtain a perfect spin filter as in
the case of noninteracting electrons. In Fig.\ \ref{fig:gH}
we plot the numerical results for the conductance obtained from
Eqs.\ (\ref{cos}) and (\ref{gsigma})
in a similar manner as in Fig.\ \ref{fig:g} 
for $U=0$ and $U=12$ with $\Delta =1.6$.

Since the dot levels are no longer degenerate the Kondo effect
is totally suppressed even at zero temperature. Indeed no 
plateaus formed near $\lt\la n_\sigma \rt\ra=1/2$ in Fig.\ \ref{fig:nH}.
As the result the spin-dependent resonances in the 
conductance have the width $\Gamma$.
The effect of large $U$ manifests itself mainly in the
mere enhancement of the level splitting.
At low temperatures $T\ll\Delta$ the mean occupation number
$\lt\la n_{\downarrow} \rt\ra$ is close to unity
while the occupation number for the opposite spin direction
decreases to zero. One can observe, however, from Fig.\ \ref{fig:nH}
that there is a mixing of the levels 
which gives rise to the non-monotonous behavior of
$\langle n_{\sigma} \rangle$. This results in a
small dip of the spin down conductance 
at the resonance energy for the spin up 
electrons and vice versa.

\section{Conclusions}

We have studied numerically 
the spin-dependent conductance of 
electrons through a single ballistic channel
weakly coupled to a quantum dot.
The Fano interference on the dot 
reverses the dependence of the channel conductance 
on the gate voltage compared to that of the dot's
conductance. If the spin degeneracy of the 
dot levels is lifted, the conductance of the channel
reveals Coulomb blockade dips, or anti-resonances,
as a function of the gate voltage.
If the level is spin degenerate the 
Kondo effect suppresses the channel conductance
in a broad range of the gate voltages.

When the coupling between the channel and the dot is not
truly local the resonances should become highly antisymmetric.
In the idealized picture of the single ballistic channel 
transport the zero temperature conductance vanishes at
the resonance values of the gate voltage due to the Fano
interference. If there is no spin degeneracy, the 
resonances have single-particle origin
(for the simple model (\ref{eq1}) 
they are separated roughly by the Coulomb energy $U+\Delta$).
At the resonance the channel conductance for a certain spin
direction changes abruptly from its ballistic value $e^2/h$
to zero, while the opposite spin conductance remains approximately
at the ballistic value. 

This is in contrast with previous  calculations that considered a substitutional 
QD, where the complementary  effect is found, i.e., a perfect conductance for one 
spin species. However, in these cases due to the local
 correlations and hibridization with the leads, there remains a finite
 conductance of the opposite spin species blurring the filtering condition. In 
the lateral dot case considered here we find {\it total} suppression of one spin 
species, thus leading to a perfect spin filter. Moreover we obtain that the 
energetical splitting of the Fano resonances for spin up and spin down electrons 
is of the order of the Coulomb interaction $U$ for large $U$ rather than of the 
order of the Zeeman splitting $\Delta$ as it would be for noninteracting 
electrons $U=0$. Thus a strong electron interaction on the dot (realistic
 values for $U$ are of the order of 1meV) will allow to apply very weak magnetic 
fields (less than 1Tesla) and still observe a significant splitting of
 the Fano resonances (provided $k_BT<\Delta$), and thus a high quality spin  
filter. This is because not only the spin filter is totally reflecting 
electrons with the desired spin $\sigma$, but it will be
 then also close to fully transparent for electrons with opposite spin.
Note, however, that the coupling to the dot
should preserve the phase coherence and there should be 
only one conducting channel.

We thank A.~Aharony, O.~Entin-Wohlman, V.~Fleurov, 
L.~Glazman, K.~Kikoin and M.~Raikh for helpful 
discussions. K.H and M.E.T acknowledge CONICET for support. 
Part of this work was done under 
projects Fundaci\'on Antorchas 14116-168 and 14116-212.


\begin{thebibliography}{10}
\bibitem{kang-remark}
For a detailed derivation in the absence of magnetic fields
we refer to Ref.~\cite{kang01}.
\end{thebibliography}
\end{document}